\documentclass{article}
\usepackage{spconf,amsmath,epsfig}
\usepackage[]{graphicx}
\graphicspath{{Figures/}}
\usepackage{caption}
\usepackage{subfig}
\usepackage{lipsum}
\usepackage{url}
\usepackage{color}
\usepackage{bm}
\usepackage{balance}
\usepackage{tabto}

\usepackage{subfig}
\usepackage{hyperref}

\usepackage{array}
\newcommand{\PreserveBackslash}[1]{\let\temp=\\#1\let\\=\temp}
\newcolumntype{C}[1]{>{\PreserveBackslash\centering}p{#1}}
\newcolumntype{R}[1]{>{\PreserveBackslash\raggedleft}p{#1}}
\newcolumntype{L}[1]{>{\PreserveBackslash\raggedright}p{#1}}

\title{On the Relationship Between Ground- and Satellite- Based \\Global Horizontal Irradiance}

\name{Mayank Jain$^{1,2^{*}}$, Deepak Joel Yericherla$^{3^{*}}$, and Soumyabrata Dev$^{1,2}$
\thanks{The ADAPT Centre for Digital Content Technology is funded under the SFI Research Centres Programme (Grant \#13/RC/2106\_P2) and is co-funded under the European Regional Development Fund.}
\thanks{Send correspondence to S.\ Dev: \protect\url{soumyabrata.dev@ucd.ie}}
\thanks{$^{^{*}}$Authors contributed equally.}
}
\address{
$^{1}$The ADAPT SFI Research Centre, Dublin, Ireland \\
$^{2}$School of Computer Science, University College Dublin, Dublin, Ireland \\
$^{3}$International Institute of Information Technology Naya Raipur, Chhattisgarh, India}
\begin{document}
%\ninept
%
\maketitle
\begin{abstract}
Global horizontal irradiance (GHI) plays a significant role in maintaining the earth's ecological balance and generating electricity in photovoltaic systems. While the satellites have more range, they have been shown to over/under-estimate the true values of GHI that are observed at the ground-based stations. Hence, this study aims at analyzing the relationship between these two sources of GHI data in order to better and effectively utilize the reach of satellites for GHI analysis. The paper identifies a near linear relationship between the two and thereby concludes that an approximate mapping from satellite- to ground-based GHI values can be obtained.
\end{abstract}
\section{Introduction}
\label{sec:intro}
Horizontal surface solar irradiance, or global horizontal irradiance (GHI), is the amount of power reaching a horizontal plane on the surface of the earth from the sun. It plays a vital role maintaining the surface energy balance and affects the behaviour and growth of flora and fauna~\cite{monteith1972solar}. It also drives various atmospheric and climate phenomenon~\cite{stephens2012update}. Apart from being essential to the very existence of life on the earth, the amount of GHI that reaches the surface of a photovoltaic system determines the amount of electrical energy that it can generate~\cite{alskaif2020systematic}. Hence a correct estimation of its value at the surface of the earth is crucial for multiple research directions.

Primarily, there are two sources for GHI data collection and estimation, i.e., satellite and ground-based sensors. Satellites can cover a larger area, including the remote locations, mountains and oceans, a feat which is not realistic with ground-based sensor systems. However, it has been noted that satellites generally provide biased estimations for the GHI values~\cite{olomiyesan2016comparative,ernst2016comparison,manara2020comparison}. Typically low spatial resolution of the satellites further elevates the problem, making their readings more inaccurate. To this end, this paper\footnote{In the spirit of reproducible research, the code related to this paper is available from  \url{https://github.com/ydjoel/SolarSatGround}.} analyzes the relationship between the satellite and ground-based sensor readings. The paper further attempts to model ground-based GHI readings from the satellite data.

\subsection{Relevant Literature}
Estimating surface solar irradiance values from the satellites has been an area of ongoing research~\cite{huang2019estimating}. Satellites typically sense the solar energy going into the top of the earth's atmosphere and the energy that is reflected back. These observations are then used to estimate atmospheric constituents and their effects on incoming solar radiations. Post accumulating these estimations the surface solar irradiance is estimated. Being such an indirect process, it becomes very difficult to correctly estimate the true ground-level values from the satellites.

In a comparative analysis, it was noted that the average errors of satellite-derived GHI readings range between $-7\%$ to upto $25\%$ in Nigeria~\cite{olomiyesan2016comparative}. Upto $9.3\%$ overestimation error in satellite-based values was reported in a separate study in Australia~\cite{ernst2016comparison}. Manara~\textit{et}~\textit{al.}~\cite{manara2020comparison} analyzed the accuracy of satellite-based GHI over varying altitude levels. It was noted that the results vary with elevation. The values were generally overestimated in low-lying areas whereas they were underestimated at more elevated locations. In general, there was a question on the accuracy and effectiveness of the methods which are being used to estimate GHI from the satellite readings.

\section{Dataset}\label{sec:dataset}
The data was separately downloaded for ground-based stations and the satellite derived readings. The details for both datasets is discussed in the following subsections.

\subsection{Ground-based Sensor Data}
We collected the GHI land-based data from the Solcast website~\cite{solcast}. Specifically, the dataset provides total irradiance or GHI that is received on a horizontal surface on the ground. It is the sum of direct and diffuse irradiance components. In this study, we choose Dublin as the city under consideration. By default, Solcast provides data from the nearest available solar farms given the latitude and longitude information.% The tools check for available solar farms or large-scale rooftop Photovoltaic sites nearby the latitude and longitude provided.

Seven years worth of data was obtained from $2014$ to $2020$. The data consists of the timestamp and the GHI readings in $W/{m}^{2}$. Apart from that, other vital information about the exact location, altitude and time zone is provided in the dataset. To organize the data better, it was separated into lists where data from the same day are kept together similarly days of a month are held together and months of the same year.
%The data can be requested for a specific time interval. We got values of GHI from 2014 to 2020, seven years. We would get the data in CSV format. It would contain the following columns;  Year, Month, Day, Hour, GHI (in W/M2 ). The file also includes vital information, such as the coordinates of the solar site we are using, country, Data Source, Timestep, Latitude, altitude, and time zone. Each data point has an associated timestamp corresponding to the Hour the measurement took. With the timestamps in extended form, the data can easily be separated into lists where data from the same day are kept together similarly days of a month are held together and months of the same year. This was done so that the data could be analyzed more easily.
\subsection{ERA5 Data}
The satellite-based solar irradiance data is compiled from the Climate Data Store (CDS) website which was provided by the European Centre for Medium-Range Weather Forecasts (ECMWF)~\cite{era5paper}. The ERA5 dataset provides hourly estimates for many atmospheric, ocean-wave, and land-surface quantities. In this case the estimated surface level GHI values were provided under the name of `Surface Solar Radiation Downwards' variable.% An underlying 10-member ensemble samples an uncertainty estimate at three-hourly intervals.

%As mentioned earlier, we are considering Dublin to carry out this analysis.
The data was obtained by making API calls to the CDS Server. A POST request is sent with exact specifications of the variable name, timestamps, and the geographical area to obtain the final response in NetCDF format. In this case, the data was downloaded for the exact same location for which the ground-based data was taken. Since the ERA5 data is the hourly reanalyses data over $3$ hours, a shift in sequential data by $2$ time-steps was required to match the timestamp of the ground-based sensor data. Furthermore, the raw satellite data was divided by $3 \times 3600$ to convert the units from $Js^{-1}m^{-2}$ to $Wm^{-2}$ to match the units that were obtained from the ground-based sensors.%the accumulation period is over one 1 hour ending at the validity date and time, so we have to shift the radiation values by a unit.%To do that, we have to set up a client instance with our key (which can be generated by signing up at CDS and accepting the terms). We have to send a POST request with a variable name ('Surface Solar Radiation Downwards' here), the year, month, day, hour, and finally, the geographical area we require the values. We would get the response in NetCDF format.

%If the coordinates are far apart, more points will be between the coordinates. To minimize that, the coordinates for each region were chosen so that only 9 data points were downloaded. The data point at the center of each area is as close as possible to the land stations so that this point will be used for comparison with the land data.

%Here, the variable used for satellite solar radiation is 'ssrd' , which stands for Surface solar radiation downwards. This parameter is the amount of solar radiation (also known as shortwave radiation) that reaches a horizontal plane at the surface of the Earth. This parameter comprises both direct and diffuse solar radiation [ era5 ]. We downloaded the data for each hour of the day from 2014-2020.

\section{Methods}\label{sec:methods}
\begin{figure}[htb]
\centering
\subfloat[Actual GHI values as obtained from ground-based sources~\cite{solcast}]{\includegraphics[width=0.99\columnwidth]{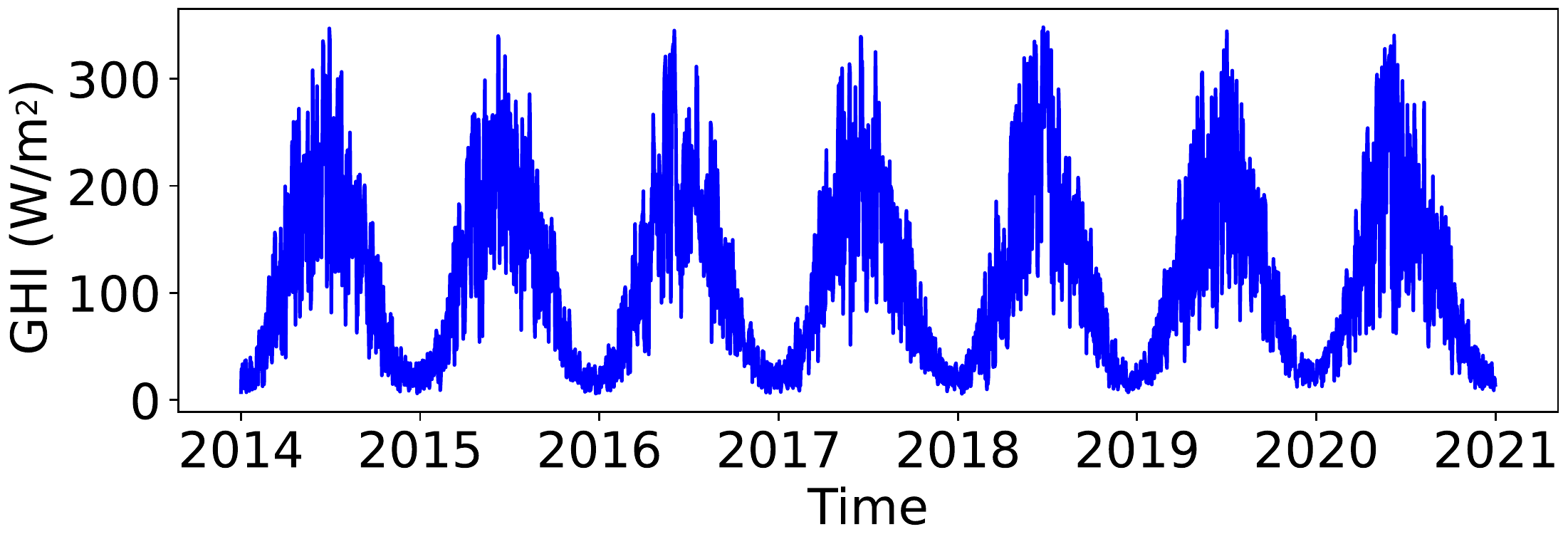}}\\
\subfloat[GHI data estimated from satellites~\cite{era5paper}]{\includegraphics[width=0.99\columnwidth]{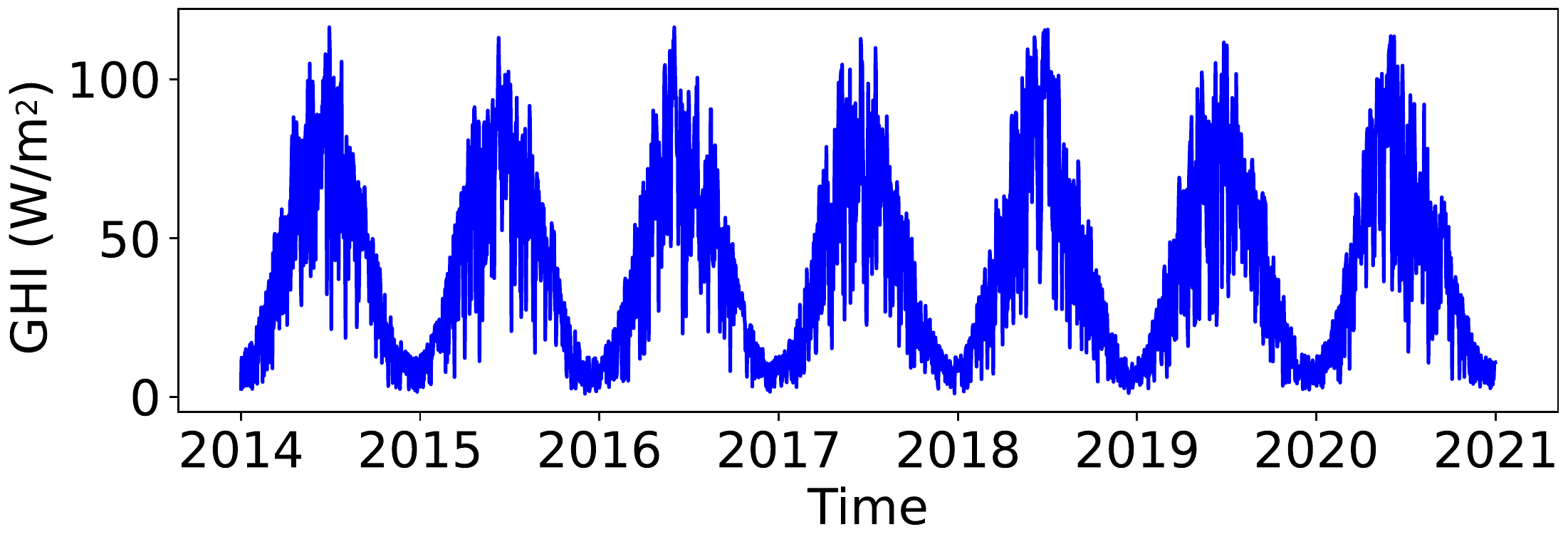}}\\
\subfloat[Difference between ground- and satellite-based observations]{\includegraphics[width=0.99\columnwidth]{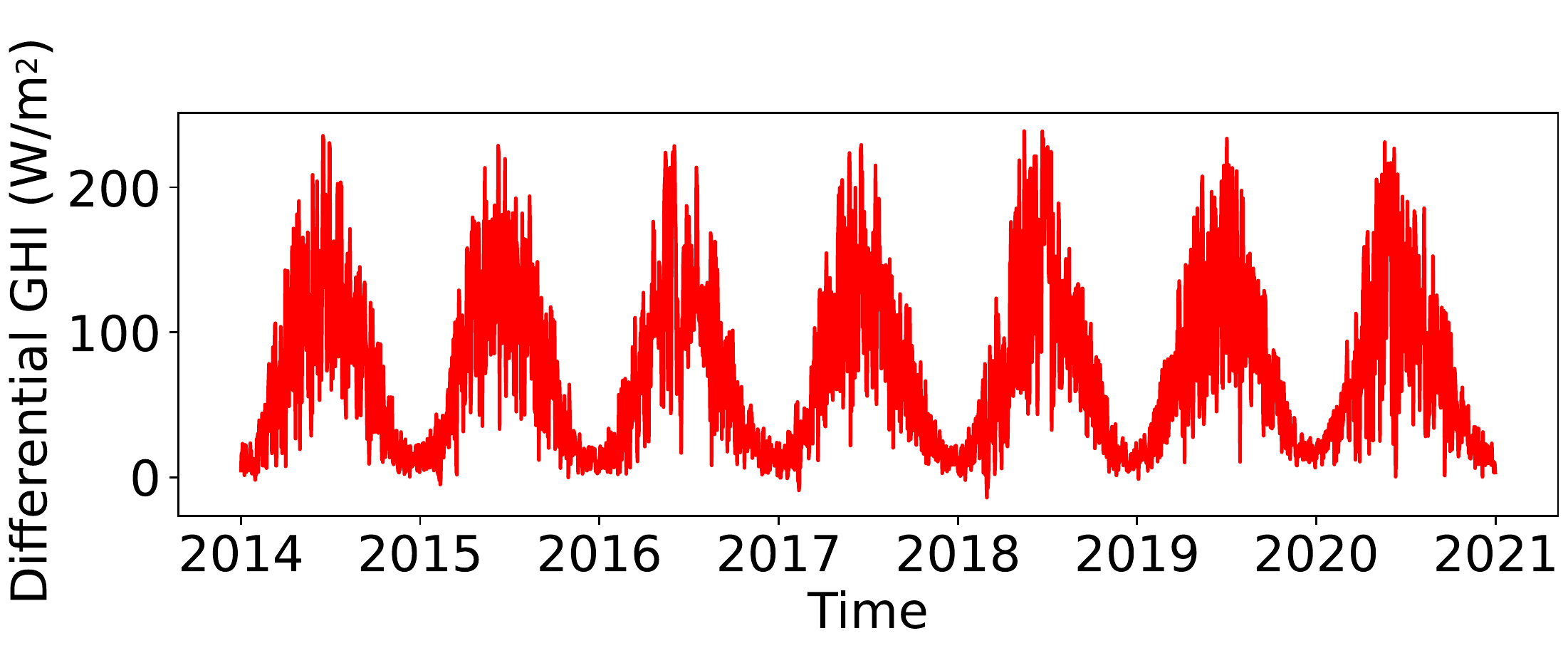}}
\caption{GHI data estimated from ground-based station and satellite data.}
\label{fig:GHI-data}
\end{figure}

Both the land and satellite datasets have solar radiation values spaced at successive intervals of an hour. Fig.~\ref{fig:GHI-data}(a) shows the actual GHI values that were obtained from the ground-based sensor dataset, whereas Fig.~\ref{fig:GHI-data}(b) shows the trend of GHI values that were estimated from the satellites.%We plotted these values to get a rough understanding of the data we are dealing with. 

% \begin{figure}[!ht]
%     \centering
%     \includegraphics[width=0.99\columnwidth]{land.pdf}
%     \caption{Actual GHI values as obtained from ground-based sources~\cite{solcast}}
%     \label{fig:groundData}
% \end{figure}

% \begin{figure}[!ht]
%     \centering
%     \includegraphics[width=0.99\columnwidth]{sat.pdf}
%     \caption{GHI data estimated from satellites~\cite{era5paper}}
%     \label{fig:satelliteData}
% \end{figure}

It can be clearly noted that the radiation is high in warmer months of a year but comes down progressively as we get to December and January. This cycle continues each year. The only notable difference between the two plots is the peak value they attain. Ground-based sensors seems to record higher observations than its satellite counterparts. In other words, satellites are generally underestimating the true GHI values in this case. To compare the difference between the readings obtained from the two datasets, a difference curve was plotted by subtracting the satellite readings from the corresponding ground-based sensor readings. The obtained difference plot in Fig.~\ref{fig:GHI-data}(c) confirms that there is considerable difference in the satellite estimations and the actual values of the GHI at the earth's surface.

%The most straightforward method of comparing the radiation values from the datasets is to subtract the land values for each day from the satellite values. The difference between the land and satellite data indicates how closely related the data sets are. 

% \begin{figure}[!ht]
%     \centering
%     \includegraphics[width=0.99\columnwidth]{diff.pdf}
%     \caption{Difference between ground- and satellite-based observations}
%     \label{fig:differencedData}
% \end{figure}

\begin{figure}[!ht]
    \centering
    \includegraphics[width=0.99\columnwidth]{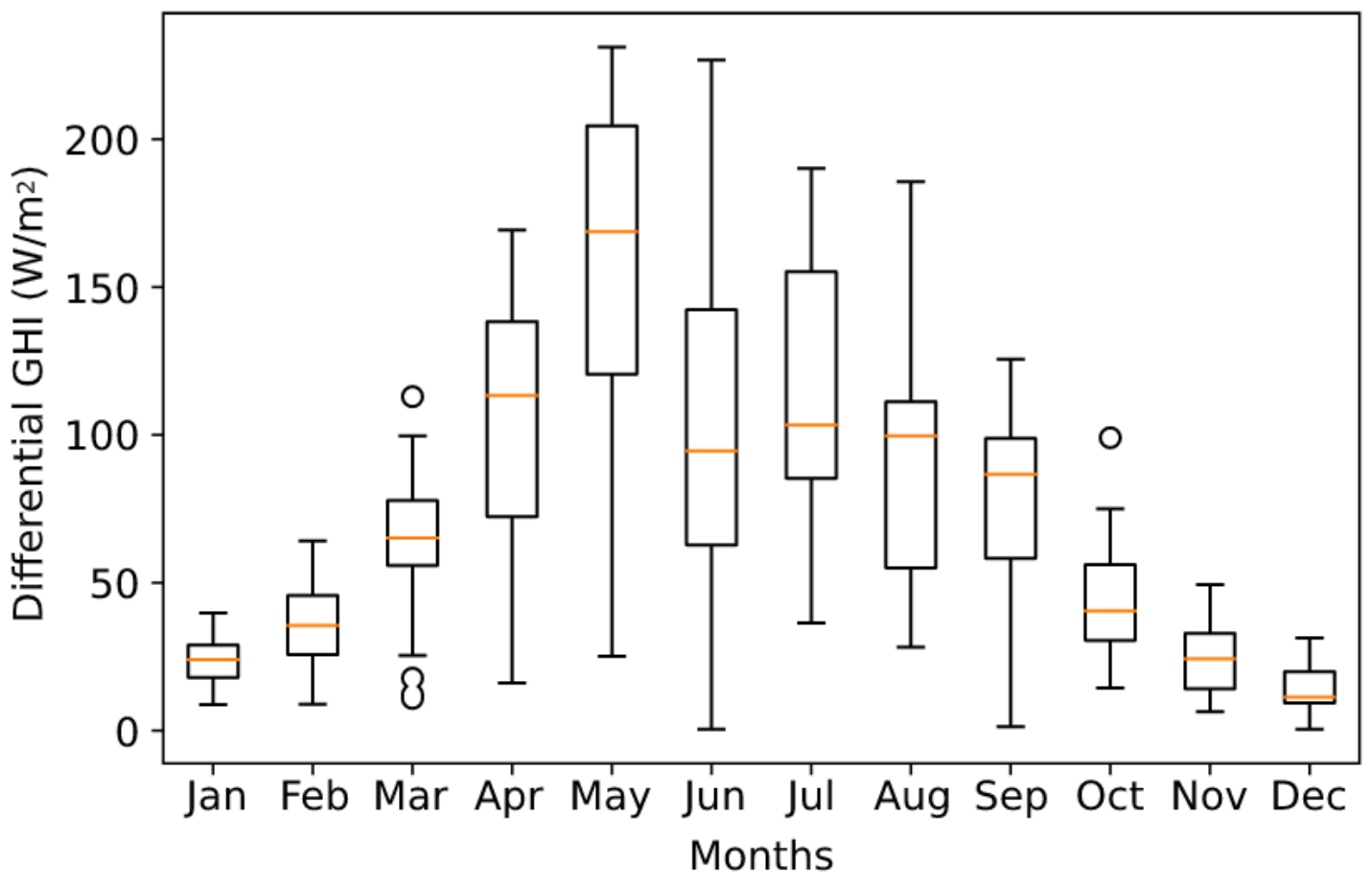}
    \caption{Daily mean difference between satellite and ground-based sensor data as obtained for different months in $2020$}
    \label{fig:boxplot2020}
\end{figure}

\begin{figure}[!ht]
    \centering
    \includegraphics[width=0.99\columnwidth]{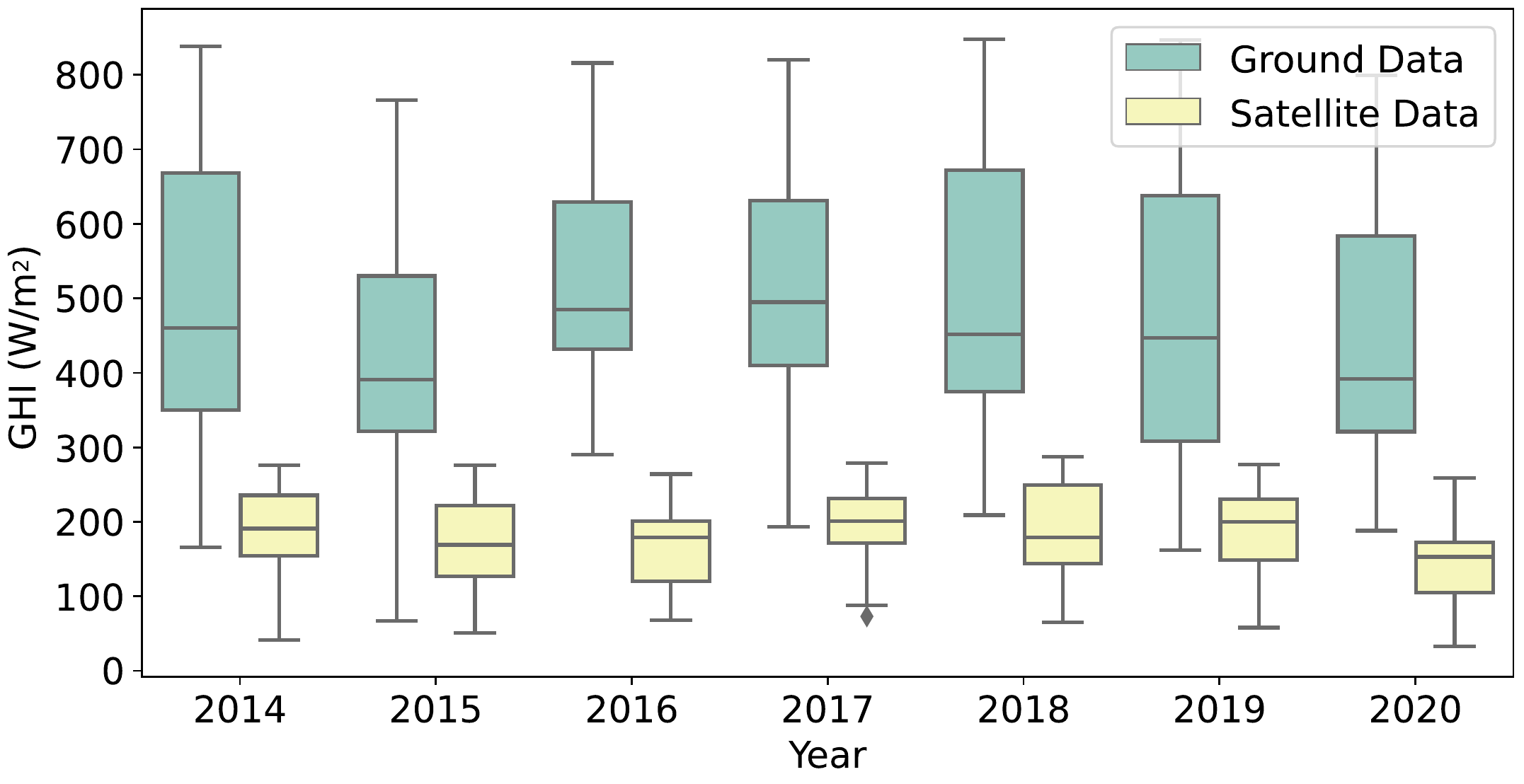}
    \caption{Variation in daily mean GHI values from ground- and satellite- based sources for the month of August from $2014$ and $2020$}
    \label{fig:augustBoxplots}
\end{figure}

To further understand the impact of temporal variations on the data, boxplots of daily mean differential GHI were plotted against different months for different years. Fig.~\ref{fig:boxplot2020} shows one such plot for the year $2020$. It can be clearly seen from the figure that there is a huge amount of variations across different months. However, on the other hand, there is no such significant difference between in the GHI values of the same month over the years. Fig.~\ref{fig:augustBoxplots} shows the boxplots of daily mean GHI values for the August month over the years. Overall, it can be noted that the variation over months is much more considerable than over the years. Hence, the paper attempts to create different models for each month for better accuracy.

%\begin{figure}[!htb]
%\centering
%\subfloat[GHI from ground-based sensors in August]{\includegraphics[width=0.99\columnwidth]{August - Land Values.pdf}}\\
%\subfloat[GHI from satellite-based sensors in August]{\includegraphics[width=0.99\columnwidth]{August - Satellite Values.pdf}}
%\caption{Variation in daily mean GHI values from (a) ground- and (b) satellite- based stations for the month of August from $2014$ and $2020$}
%\label{fig:augustBoxplots}
%\end{figure}

%\begin{figure}[!ht]
%    \centering
%    \includegraphics[width=0.99\columnwidth]{August - Land Values.pdf}
%    \caption{GHI Values from Ground-based stations for August Month from 2014 to 2020}
%    \label{fig:augustGround}
%\end{figure}

%\begin{figure}[!ht]
%    \centering
%    \includegraphics[width=0.99\columnwidth]{August - Satellite Values.pdf}
%    \caption{GHI Values estimated from satellite data for August Month from 2014 to 2020}
%    \label{fig:augustSatellite}
%\end{figure}

\begin{figure}[!ht]
     \centering
     \includegraphics[width=0.99\columnwidth]{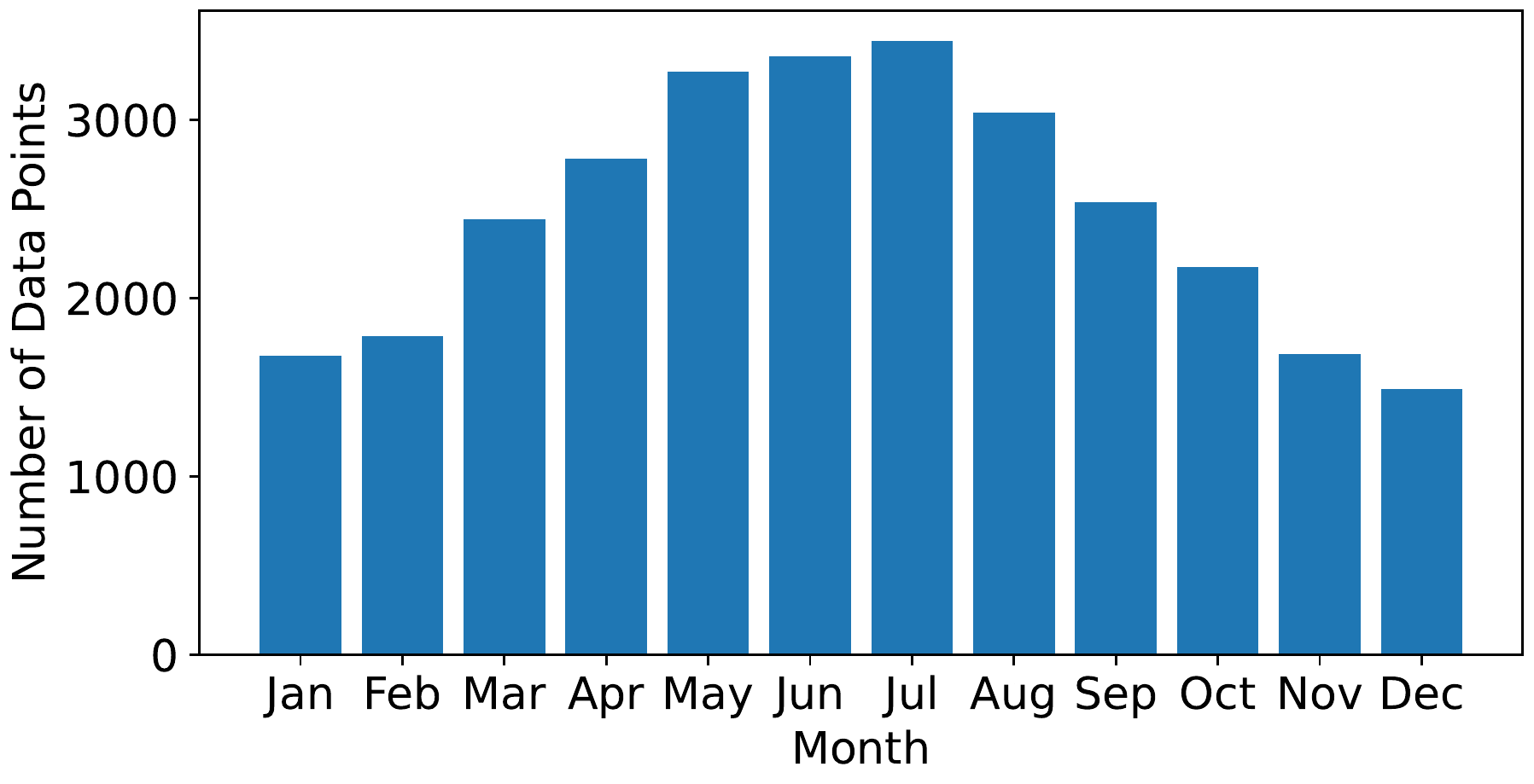}
     \caption{Number of daytime data points that were available for each month accummulated from $2014-2020$}
     \label{fig:dataPoints}
\end{figure}

While the GHI values are reported for the whole $24$ hours in both the datasets, they are $0$ (or nearly $0$) at nighttime. The length of nighttime also varies across the year as nights are longer in winters but much shorter in summers. However, in any case, all such timestamps were removed where either of the ground- or satellite-based readings were $0$. Fig.~\ref{fig:dataPoints} show the number of remaining data points (across the years) that were considered for further analysis post this stage. A clear bell-shaped curve can be seen re-emphasising the idea that summers have longer days than winters. %Both Satellite and Land data have values starting from 0 hours to 23 hours. But radiation in terms of GHI is only significant from around 7'o clock. It is also subjective on the time of the year, i.e., if it's winter, it might be a bit late than that, and the number of hours of significant radiation might also fall when we look at various plots we plotted so far, we see that is the case.

%To get a real sense of the data, we are only considering timestamps with Non zero values. We made a list of Zero Value indexes and removed them from both datasets; this was done twice, one where we got Zero indices from Land data and removed those from both data sets, reindex them and do this again for Satellite data. The results can be seen in the figure below.

%As hypothesized, there are fewer non-negative values in January, and as the months progressed, we can see the count increasing until July and falling again gradually.

Once the data has been filtered for relevant values, it is important to identify underlying patterns in the data to establish a mapping from satellite-based readings to true ground-based GHI values. For this case, linear regression was performed. As noted before, for better analysis, individual models must be trained for each month. Consequently, the data was further divided into $12$ parts by months over which independent linear regression models were trained. Note that the dataset was combined over the years as no significant variation across the years was noticed.

Since a significant variation in GHI values can be noted across the day, it is important to incorporate the timestamp as input feature to the regression models. Individual components of timestamps (i.e. day of month and hour of the day) were extracted and converted into one-hot encoded vectors. These were then concatenated to result in the final input feature vector. Since different models were created for the different months and each month has similar number of daylight hours, the size of one-hot encoded `hour' vectors will vary from one month to another as per the number of daylight hours in that particular month. Same goes for the `days' vector as well. Thus, including the satellite derived GHI readings, atleast $40$ features were created for a particular month. Finally each month's data was randomly shuffled and an $80$-$20$ split was made to divide the data into training and test set respectively. Coefficient of determination ($R^{2}$) was used to evaluate the model's performance.

%We wanted not to train a single model for all the data but individual models for each month. A single model can't be an accurate estimate since the variability of each month is different. We can also get important information from what day it is and what hour of the day it is. Every 15th day of the month might be the tipping point of decline.

%Hence, We create dummy variables for each  Day, Hour resulting in at least 40 columns a month from the existing 3 Variables ( Day, Hour, GHI) in a month. To do the analysis, we split each month's data into training and testing datasets; we took 20% of the whole set for the test dataset, and the split is done at random. 

%Consequently, a linear regression model was trained for every month individually to finally create a mapping from the satellite data to the actual ground-based sensor values.

\section{Results and Discussions}\label{sec:results}

\begin{figure*}[htb]
\centering
\subfloat[Regression over the month of January]{\includegraphics[width=0.99\columnwidth]{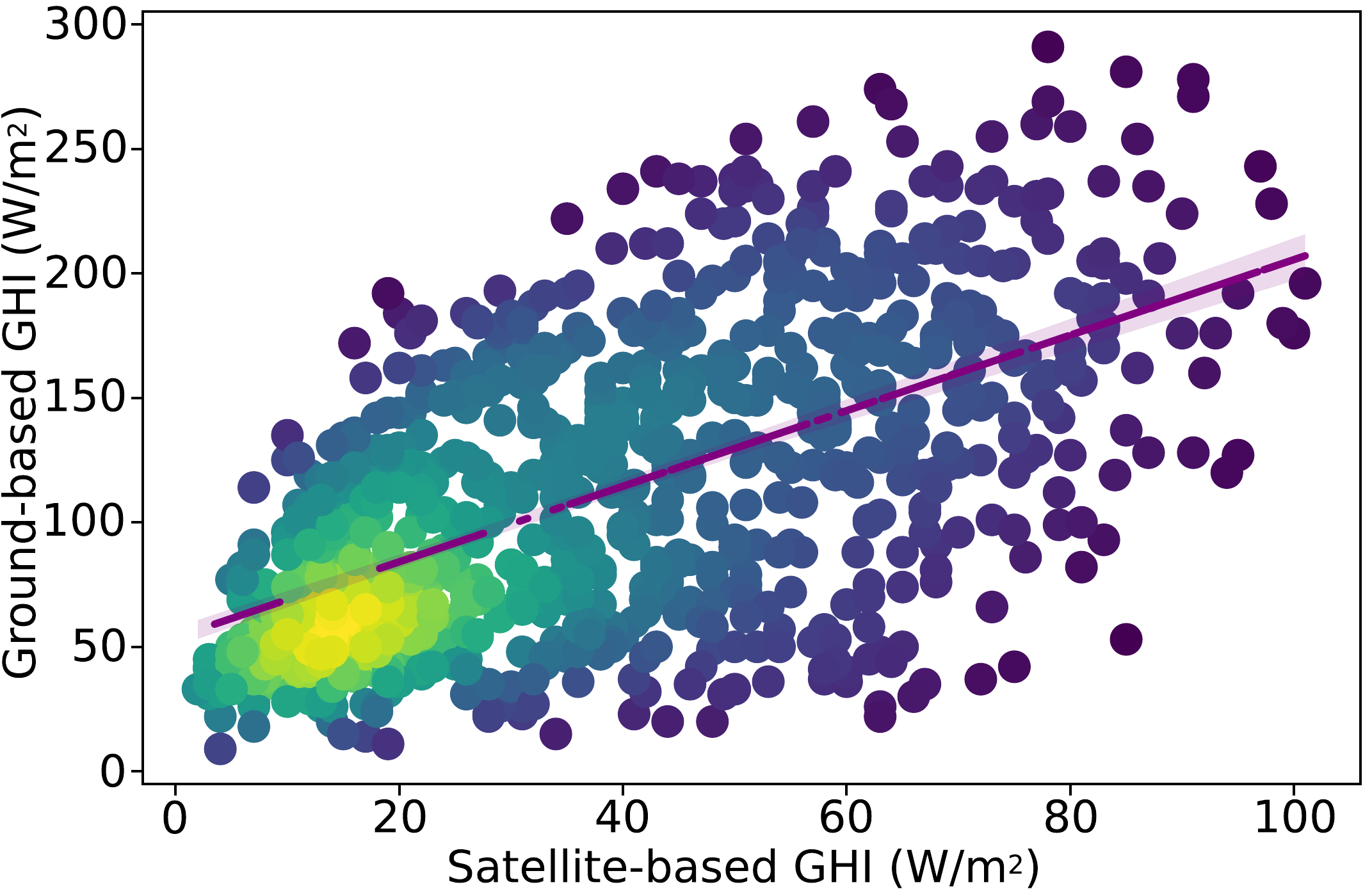}}\hfill
\subfloat[Regression over the month of June]{\includegraphics[width=0.99\columnwidth]{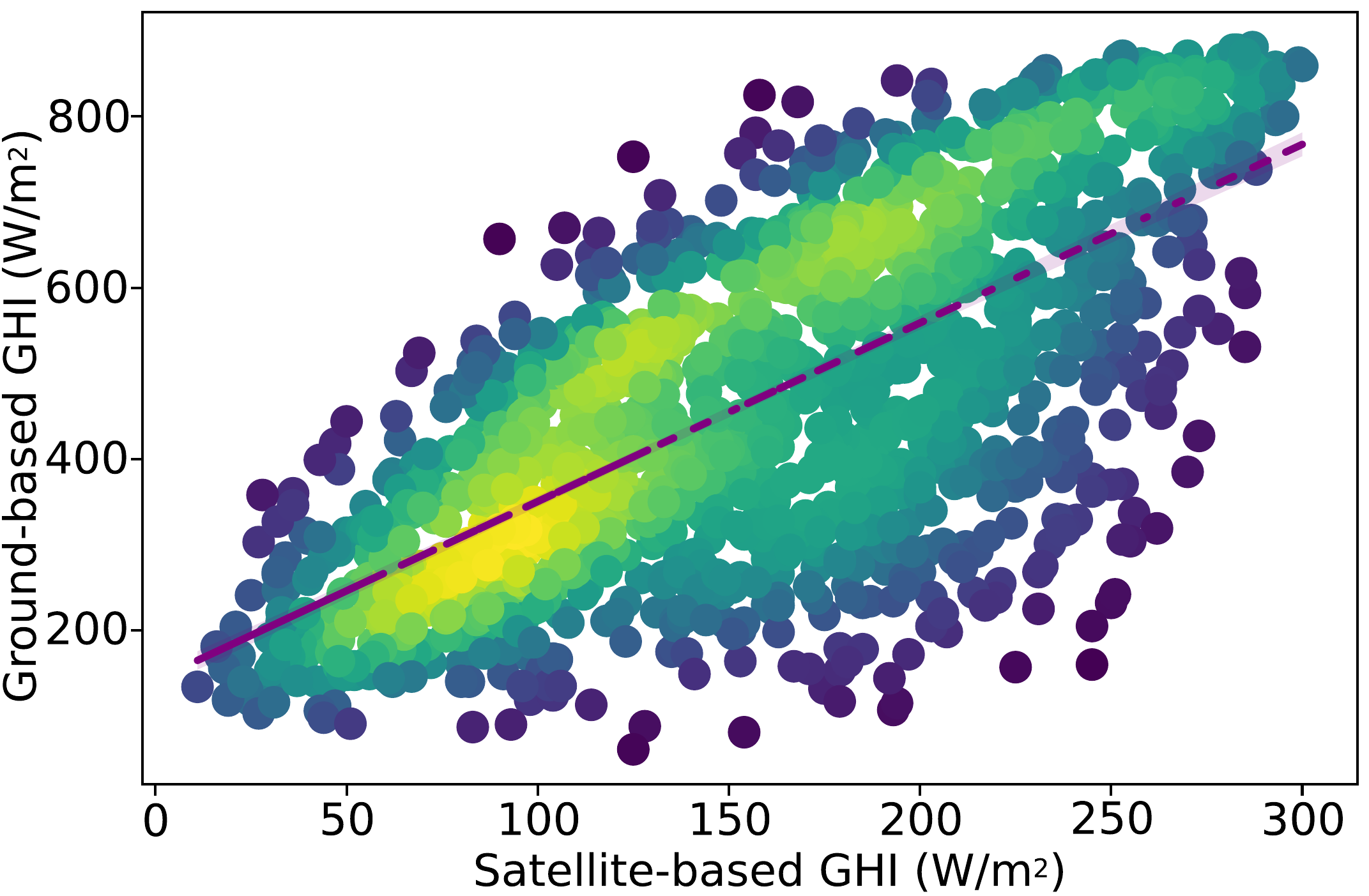}}
\caption{Scatter plot showing the satellite readings and the corresponding ground-based sensor readings in the test set. The line through the center shows the linear regression fit on the data.}
\label{fig:linRegression}
\end{figure*}

Fig.~\ref{fig:linRegression}(a) and~\ref{fig:linRegression}(b) shows the linear regression results on the test set for two sample months of January and June, respectively. The data can surely be seen to be highly correlated and a simple linear regression itself decently approximates the underlying data. However, its still not accurate by a long shot.

\begin{figure}[!ht]
    \centering
    \includegraphics[width=0.99\columnwidth]{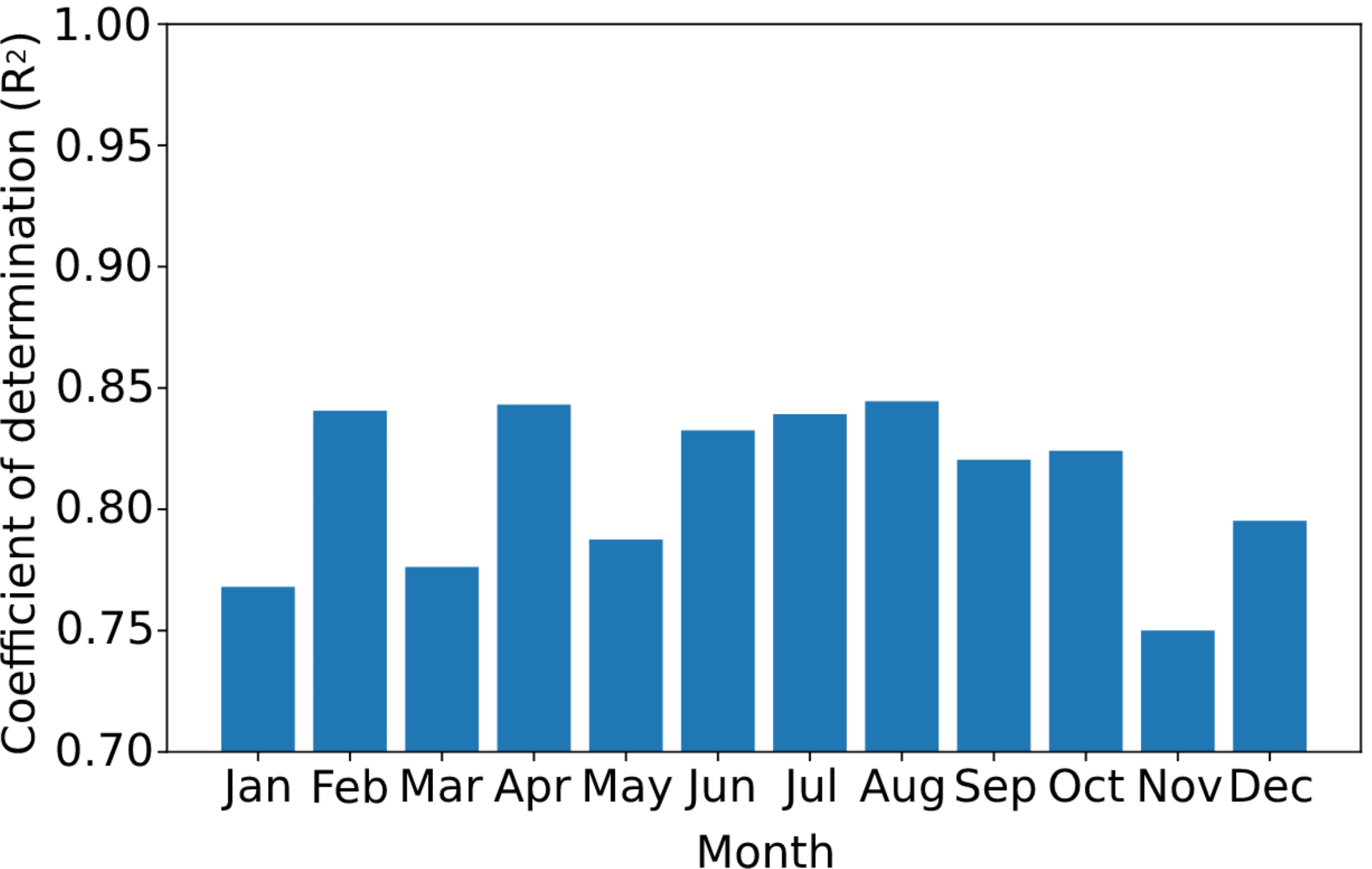}
    \caption{Coefficient of determination Values for each month}
    \label{fig:r2bar}
\end{figure}

To evaluate the fit coefficient of determination ($R^{2}$) metric was used. It gives a goodness-of-fit measure for linear regression model. $R^{2}$ explains how much of the variance in dependent variable can the independent variables collectively explain. Fig.~\ref{fig:r2bar} shows the obtained $R^{2}$ values for all $12$ models, where each of them correspond to the respective month in the year. As seen from the figure, the value of $R^{2}$ falls between $0.75$ and $0.85$. This indicates that although linear fit is definitely not the best approximation, it certainly proves that such a mapping is possible with more complex models like neural networks.

%\[R^2 = 1 - \frac{SS_{res}}{SS_{tot}}\]
%\newenvironment{conditions}
%  {\par\vspace{\abovedisplayskip}\noindent\begin{tabular}{>{$}l<{$} @{${}={}$} l}}
%  {\end{tabular}\par\vspace{\belowdisplayskip}}

%where:
%\begin{conditions}
% R^2      &  coefficient of determination \\
% SS_{res} &  sum of squares of residuals \\   
% SS_{tot} &  total sum of squares
%\end{conditions}

%\[R^2 = 1 - \frac{\Sigma(y_i - \hat{y_i})^2}{\Sigma(y_i - \hat{y})^2}\]

%We have fit a linear regression model; we need to determine how well the model fits the data. We know that the model tries to identify the equation, which results in the smallest possible squared aggregate difference between dependent and independent variables. To evaluate the fit we will use Coefficient of determination R-Squared metric. It gives a goodness-of-fit measure for linear regression model. 

%R-sqaured explains how much of the variance in dependent variable can the independent variables collectively explain. It is a percentage value. With this we can gaguge the strength of our model with respect to dependent variable

%Dependent Variable = (Constant +Independent Variables) + Error

\section{Conclusion \& Future Work}\label{sec:conclusion}
The paper presents a systematic analysis of ground- and satellite- based datasets of global horizontal solar irradiance (GHI). It was noted that satellite estimations are generally significantly off than the true ground-level observations. Not only that, but this disparity varies significantly across the months of the year. As such the paper recommends creating different models for different months of the year in order to find the best mappings from the satellite data to true GHI values. Lastly, it was shown that this mapping is almost linear but a significantly better fit might be obtained by using more complex models than linear regression. In future, the authors would like to analyze the relationship even further and try to model it with better coefficient of determination scores. Additionally, the plan is to study the generalizability of the identified models and/or the approach that is discussed in the paper for different locations on the earth.

% References should be produced using the bibtex program from suitable
% BiBTeX files (here: strings, refs, manuals). The IEEEbib.bst bibliography
% style file from IEEE produces unsorted bibliography list.
% -------------------------------------------------------------------------

\balance
% Generated by IEEEtran.bst, version: 1.14 (2015/08/26)

\end{document}